\documentclass[aps,preprint]{revtex4}
\usepackage{amsmath}
\usepackage{bm}
\usepackage{srcltx}
\usepackage{graphicx}

\graphicspath{{../pics/}}

\newcommand{\ve}{\mathbf{e}}
\newcommand{\vf}{\mathbf{f}}

\newcommand{\vvr}{\mathbf{r}}

\newcommand{\vcl}{\mathbf{L}}
\newcommand{\vcr}{\mathbf{R}}

\newcommand{\brho}{\bm\rho}

\begin{document}

\title{Eckart frame Hamiltonians in the three-body problem}

\author{A.~V.~Meremianin}
\email{meremianin@phys.vsu.ru}
\affiliation{Department of General Physics,
  Voronezh State University, 394006, Voronezh, Russia}

\date{\today}

\begin{abstract}
  The Eckart frame is used to separate out the collective rotations
  in the quantum three-body problem.
  Explicit expressions for the corresponding rotational and vibro-rotational
  (i.e. Coriolis) Hamiltonians are derived.
  Special attention is paid to the situation when two principal moments of
  inertia are equal in the equilibrium configuration. 
\end{abstract}

\maketitle

\section{Introduction}
\label{sec:intro}

The theoretical study of various properties of molecules and atomic nuclei
requires the separation of their motion into the vibrational and rotational
parts.
Such a separation can be achieved by introducing the rotating reference
frame (the body-frame, BF) whose axes are defined by the orientation of the
whole system in space.
As a result, the motion of particles with respect to BF can be considered as
``vibration'' while the motion of BF itself can be understood as the overall
``rotation''.
At this stage the question occurs as to how to choose BF in an
optimal way\footnote{Since the choice of BF is not unique, its definition can
  be considered as the gauge convention \cite{littlejohn97:_gauge}.}.
An optimal BF would be that in which the couplings between the vibrations and
the overall rotation are minimal.

C. Eckart introduced BF \cite{eckart35:_eckart_frame} which minimizes 
the vibro-rotational (or, Coriolis) couplings in case when particles of the
system perform small vibrations close to some equilibrium configuration.
The Eckart BF is that in which the following relation is fulfilled 
\begin{equation}
  \label{eq:econd-1}
  m_1  \, [\vcr_1 \times \vcr_1^{(eq)}]
+ m_2  \, [\vcr_2 \times \vcr_2^{(eq)}]
+ m_3  \, [\vcr_3 \times \vcr_3^{(eq)}] =0,
\end{equation}
where $\vcr^{(eq)}_i$ is the equilibrium value of the position vector $\vcr_i$
of the particle with the mass $m_i$.
The above equation is usually called ``the second Eckart condition''
\cite{eckart35:_eckart_frame,biedenharn}.
(The first Eckart condition is the requirement of the origin of BF to be at
c.m. of the system.)

The Eckart BF has been studied for a long time
\cite{watson68:_mol_ham,dymarsky:124103,%
mardis97:_eckart_hamilt,natanson89:_eckart_frame_3bd,%
natanson85:_planar_eckart,adler-golden85:_eckar_sym_3at_molec,%
redding79:_eckar_euler_angles,jorgensen78:_eckar_lowdin_canon_orthonorm,%
meyer78:_molec_symm_eckar_frame,biedenharn76:_nucl_vibrot_w_vortex,%
louck76:_rmp_eckar,kiselev68:_opt_spectr_eckart_el_nucl_molec,ferigle:102,%
wei03:_eckar,wei03:_eckar_4part_planar,carrington98:_eckart_chem_plett,%
carrington97:_eckart_jcp_again,tucker97:eckar_jacob_radau}.
The reason is that the vibro-rotational decomposition of the Hamiltonian
corresponding to Eckart BF has some attractive features.
One of which is that the Coriolis Hamiltonian in the Eckart frame is small for
small-amplitude vibrations.
Despite the broad literature on the topic some problems still remain.
Namely, existing results related to the three-body problem are rather
cumbersome.
They also do not exhibit simple behavior in the rigid-body limit when
vibrations tend to zero.
In this limit the Coriolis (i.e. vibro-rotational) part of Eckart frame
Hamiltonian vanishes and the rotational part of the kinetic energy becomes
that of a rigid body.

The goal of this article is to present the vibro-rotational Hamiltonians of
the Eckart frame in most simple and compact form such that their rigid-body
limit would be obvious. The consideration is based on the general formalism
developed in \cite{meremianin03:_phys_rep,meremianin04:_eckart}.

The procedure of the separation of rotations and vibrations consists in
the transformation of the total Hamiltonian into three parts
\begin{equation}
  \label{eq:ham-1}
  H = H_0 + H_{cor} + H_{rot},
\end{equation}
where $H_0$ is the vibrational Hamiltonian, $H_{rot}$ is the rotational
Hamiltonian which is quadratic in the components of the total angular momentum
$\mathbf{L}$ given in BF, and $H_{cor}$ is the vibro-rotational (Coriolis)
Hamiltonian which is linear both in components of $\vcl$ and the vibrational
linear momenta.


\section{The kinetic energy and Eckart condition
  in terms of Jacobi vectors}
\label{sec:keo-jacobi}

\begin{figure}[ptbh]
\centering
\includegraphics[width=6cm]{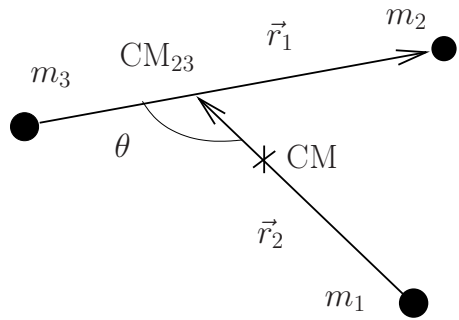}
\caption{Jacobi vectors for the three-body system. $\mathrm{CM}_{23}$ is the
  CM of the particles $m_{2}$ and $m_{3}$.}%
\label{fig:3bd-jacobi}%
\end{figure}

Assuming that the center of mass of the system is at the origin of the
coordinate frame
the Schr\"odinger equation for the system of three particles can be
written as \cite{smirnov_shitikova77:_k_harm,littlejohn97:_gauge}
\begin{equation}
\label{eq:schrod-3bd}
\left( - \frac{1}{2\mu_1}
\frac{\partial^{2}}{\partial \mathbf{r}_{1}^{2}}
- \frac{1}{2 \mu_2}
\frac{\partial^{2}}{\partial \mathbf{r}_{2}^{2}}
+U-E \right)  \Psi(\vvr_{1},\vvr_{2})=0.
\end{equation}
where $U$ is the potential energy operator which depends on three internal
variables; $\vvr_{1}$ and $\vvr_2$ are Jacobi vectors (see
fig.~\ref{fig:3bd-jacobi}) and the reduced masses $\mu_1,\mu_2$ are defined by
\begin{equation}
\frac{1}{\mu_{1}}=\frac{1}{m_{2}}+\frac{1}{m_{3}},\quad\frac{1}{\mu_{2}}
  =\frac{1}{m_{1}}+\frac{1}{m_{2}+m_{3}}.
\label{eq:def-mu12}%
\end{equation}

The second Eckart condition in terms of Jacobi vectors reads 
\begin{equation}
  \label{eq:econd-jacobi}
  \mu_1 \, [\brho_1 \times \vvr_1] 
+ \mu_2 \, [\brho_2 \times \vvr_2] =0.
\end{equation}

It is convenient to introduce the so-called ``mass-scaled'' Jacobi vectors
by replacing
$\vvr_1 \to \vvr_1/ \sqrt{\mu_1}$, $\vvr_2 \to \vvr_2 / \sqrt{\mu_2}$
(and analogously for $\brho_{1,2}$).

\begin{figure}
  \centering
  \includegraphics[width=6cm]{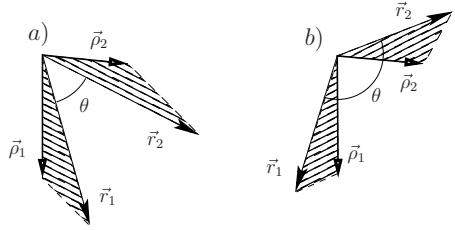}
  \caption{The geometrical meaning of the Eckart condition.
    $a)$ the mutual angle between Jacobi vectors is $\theta < \pi/2$,
    $b)$ the mutual angle is $\theta > \pi/2$.
    On each figure the areas of the dashed triangles are equal.}
  \label{fig:eckart-areas}
\end{figure}

In terms of mass-scaled Jacobi vectors the Schr\"odinger equation becomes
\begin{equation}
  \label{eq:si-mass-jac}
   \left( - \frac{1}{2} (\Delta^{2}_1 + \Delta^{2}_2)
+U-E \right)  \Psi(\vvr_{1},\vvr_{2})=0, 
\end{equation}
and the second Eckart condition reads
\begin{equation}
  \label{eq:eckond-mass-jac}
 [\brho_1 \times \vvr_1] + [\brho_2 \times \vvr_2] =0.
\end{equation}
This condition has simple geometrical meaning.
Namely, according to (\ref{eq:eckond-mass-jac}), the areas of two triangles
build on pairs of vectors $\brho_1, \vvr_1$ and $\brho_2, \vvr_2$ must be equal,
see Fig.~\ref{fig:eckart-areas}. 

The equilibrium vectors $\brho_{1,2}$ are not moving in BF and they define the
basis vectors of BF.
Let us assume that BF basis vectors are directed along the main axes of
the inertia tensor in the equilibrium configuration.
The decomposition of the vectors $\brho_{1,2}$ over the BF basis vectors
$\ve_1$, $\ve_2$ can be written as
\begin{equation}
  \label{eq:brho-comp}
  \begin{split}
    \brho_{1} &= x_1 \, \ve_1 + y_1 \, \ve_2, \\
    \brho_{2} &= x_2 \, \ve_1 + y_2 \, \ve_2,
  \end{split}
\end{equation}
where $(x_k, y_k)$ are Cartesian components of $\brho_k$ ($n=1,2$) in BF.
Now let us introduce Eckart vectors $\vf_1$ and $\vf_2$ by
\begin{equation}
  \label{eq:def-f-vect}
  \begin{split}
      \vf_1 & = x_1 \, \vvr_1 + x_2 \, \vvr_2, \\
      \vf_2 & = y_1 \, \vvr_1 + y_2 \, \vvr_2.
  \end{split}
\end{equation}
Inserting (\ref{eq:brho-comp}) into (\ref{eq:eckond-mass-jac}) one arrives
at the following representation of the Eckart condition
\begin{equation}
  \label{eq:econd-2}
  [\ve_1 \times \vf_1] + [\ve_2 \times \vf_2] =0.
\end{equation}
The solution of this equation has been obtained in
\cite{meremianin04:_eckart} and it reads 
\begin{equation}
  \label{eq:3bd-e-f}
  \begin{split}
    \ve_1 =& \frac{1}{\mathcal{F}}
    \left(
      \vf_1 + \frac{[\vf_2 \times [\vf_1 \times \vf_2]]}{|\vf_1 \times \vf_2|}
    \right), \\
    \ve_2 =& \frac{1}{\mathcal{F}}
    \left(
      \vf_2 - \frac{[\vf_1 \times [\vf_1 \times \vf_2]]}{|\vf_1 \times \vf_2|}
    \right),
  \end{split}
\end{equation}
where the Eckart parameter ${\cal F}$ is defined by
\begin{equation}
  \label{eq:2d-trace-f}
  \mathcal{F} = \sqrt{f_1^2 + f_2^2 + 2\, | \vf_1 \times \vf_2 |}.
\end{equation}
The orthogonality and normalization of vectors $\ve_1$ and $\ve_2$ as well
as their compliance with the Eckart condition (\ref{eq:econd-2})
can be easily proved by straightforward computations.

The Eckart parameter ${\cal F}$ can also be written as
\cite{meremianin04:_eckart},
\begin{equation}
  \label{eq:f-alt}
  {\cal F} = (\ve_1 \cdot \vf_1 ) + (\ve_2 \cdot \vf_2). 
\end{equation}
This equation clarifies the meaning of the Eckart parameter.
As is seen, ${\cal F}$ is the sum of projections of Eckart vectors on the
corresponding basis vectors of Eckart BF.


\section{Eckart internal variables}
\label{sec:internal-coordinates}

We choose the set of internal variables to be the lengths of the Eckart
vectors $f_1, f_2$ and cosine of their mutual angle 
$\tau = \cos \phi = (\vf_1 \cdot \vf_2)/(f_1 f_2)$.
In order of derive the expression for the kinetic energy operator we note the
connection between the gradient operators in terms of Jacobi and Eckart
vectors,
\begin{equation}
  \label{eq:j-eck}
  \frac{\partial}{\partial \vvr_\alpha} = 
x_\alpha \, \frac{\partial}{\partial \vf_1}
 + y_\alpha \, \frac{\partial}{\partial \vf_2}, 
\quad \alpha =1,2.
\end{equation}
From this equation one can derive the expression for the kinetic energy,
\begin{equation}
  \label{eq:keo-f}
  \nabla^{2}_1 + \nabla^{2}_2
= I_2 \, \tilde\nabla^{2}_1 + I_1 \,  \tilde\nabla^{2}_2,
\end{equation}
where $\tilde\nabla_{1,2}$ is the gradient operator with respect to
$\vf_{1,2}$ and $I_2$ ($I_1$) is the equilibrium moment of inertia with respect to 
$y$- ($x$-) axis (see Appendix \ref{sec:equil-I}),
\begin{equation}
  \label{eq:def-I}
  I_1 = y_1^{2}+y_2^{2}, \quad
  I_2 = x_1^{2}+x_2^{2}.
\end{equation}
When deriving (\ref{eq:keo-f}) we have used the fact that the axes of BF are
directed along the principal axes of the equilibrium inertia tensor, so that
\begin{equation}
  \label{eq:bf-2}
  x_1 y_1 + x_2 y_2 =0.
\end{equation}

The connection of the Eckart internal variables with the Jacobi-bond variables
follows from the definition (\ref{eq:def-f-vect}) of vectors $\vf_{1,2}$ and
is given in Appendix~\ref{sec:conn-eckart-jacobi}.

The above eq.~(\ref{eq:keo-f}) allows one to derive the expression for the
vibrational kinetic energy in terms of variables $f_1,f_2,\tau$.
Omitting details of some routine computations which are analogous to those
presented in \cite{meremianin03:_phys_rep} the final result reads
\begin{equation}
  \label{eq:h-vib}
  H_0 = -\sum_{i=1}^{2} \frac{I_i}{2 f_i^{2}} \frac{\partial}{\partial f_i}
f_i^{2} \frac{\partial}{\partial f_i}
- \frac{1}{2}
\left( 
 \frac{I_1}{f_1^{2}} + \frac{I_2}{f_2^{2}} 
 \right)
\frac{\partial}{\partial \tau} (1-\tau^{2}) \frac{\partial}{\partial \tau}
+ U (f_1, f_2, \tau),
\end{equation}
where $U$ denotes the potential energy which depends only on the internal
variables.


\section{The Coriolis Hamiltonian}
\label{sec:coriolis-hamiltonian}

In \cite{meremianin03:_phys_rep,meremianin04:_eckart} the general expression
for the Coriolis Hamiltonian has been found which is
\begin{equation}
  \label{eq:h-cor-1}
      H_{cor} = i \sum_{k=1}^3  L_k 
  \left(
  \sum_{\gamma=1}^{3} 
    C_{\gamma k} (\xi) \frac{\partial}{\partial \xi_\gamma}
  + B_k (\xi)
  \right).
\end{equation}
Here, $L_k$ denotes the projection of the total angular momentum operator on
the $k$-th axis of BF ($k=1,2,3$) and
the parameters $C_{\gamma k}$ and $B_k$ are determined by the differential
properties of BF basis vectors.
For the Eckart BF these parameters were calculated in
\cite{meremianin04:_eckart}.

In the three-body problem among parameters $B_k$ only $B_3$ is non-zero
\cite{meremianin04:_eckart},
\begin{equation}
  \label{eq:b-z}
  B_3 = \frac{I_2 - I_1}{\mathcal{F}^2} \,
    \frac{(\vf_1 \cdot \vf_2)}{|\vf_1 \times \vf_2|}
= (I_2 - I_1)\, \frac{\cot \phi}{{\cal F}^{2}},
\end{equation}
where the Eckart parameter ${\cal F}$ is defined by (\ref{eq:2d-trace-f}).

The parameters $C_{\gamma k}$  depend on the choice of the internal variables.
The expression for $C_{\gamma k}$ is given in \cite{meremianin04:_eckart},
\begin{equation}
  \label{eq:c-gk}
    C_{\gamma k}
= - \sum_{ijq}
(\mathbf{U}^{-1})_{kq } \, \epsilon_{ij q} \, 
\sum_{\alpha} \eta_{\alpha i}\,
(\ve_j \cdot \nabla_\alpha) \, \xi_\gamma 
= - \sum_{q}
(\mathbf{U}^{-1})_{kq }  
\sum_{\alpha=1}^{2}
(\ve_q \cdot [\brho_\alpha \times \nabla_\alpha]) \, \xi_\gamma 
\end{equation}
Here, $\epsilon_{ijq}$ is the unit totally antisymmetric tensor, 
$\mathbf{U}^{-1}$ is a $3 \times 3$-matrix whose elements depend on $\xi$.
In the three-body problem explicit form of $\mathbf{U}^{-1}$ can be obtained
using results of \cite{meremianin04:_eckart},
\begin{equation}
  \label{eq:u-inv}
\mathbf{U}^{-1} = \frac{1}{{\cal F}\, | \vf_1 \times \vf_2|}
\left(
  \begin{array}{ccc}
    | \vf_1 \times \vf_2| + f_1^{2} & (\vf_1 \cdot \vf_2) & 0 \\
    (\vf_1 \cdot \vf_2) & | \vf_1 \times \vf_2| + f_2^{2} & 0 \\
   0 & 0 & | \vf_1 \times \vf_2|
  \end{array}
\right).
\end{equation}

Noting the definition (\ref{eq:def-f-vect}) of Eckart vectors,
expression (\ref{eq:j-eck}) for the gradient operators $\nabla_\alpha$ 
and the property (\ref{eq:bf-2}) one can prove that the equation
\begin{equation}
  \label{eq:nabla-f}
\sum_{\alpha=1}^{2} [\brho_\alpha \times \nabla_\alpha]
= I_2 \, [\ve_1 \times \tilde\nabla_1]
+ I_1 \, [\ve_2 \times \tilde\nabla_2]
\end{equation}
is valid.
The action of the operators $\tilde\nabla$ on the internal variables $\xi$
gives a vector lying in the $xy$-plane of BF.
This fact, together with (\ref{eq:u-inv}), leads to the conclusion that among
coefficients $C_{\gamma k}$ only $C_{\gamma 3}$ is non-zero. It is
\begin{equation}
  \label{eq:c-g3}
  C_{\gamma 3} = -\frac{1}{{\cal F}} 
\bigl( I_2 (\ve_2 \cdot \tilde\nabla_1) - I_1 (\ve_1 \cdot \tilde\nabla_2)
\bigr)\, \xi_\gamma.
\end{equation}
This equation is the main result of the present section.

The set of $C$-coefficients corresponding to Eckart internal
variables $\xi = (f_1 , f_2, \cos \phi)$ defined in
Sec.~\ref{sec:internal-coordinates} can easily be obtained using
(\ref{eq:c-g3}). The result reads
\begin{equation}
  \label{eq:c-3-xi}
  \begin{split}
    C_{1 3} &= - \frac{I_2}{{\cal F} f_1 } (\ve_1 \cdot \vf_1)
= - I_2 \frac{ (\vf_1 \cdot \vf_2 ) }{{\cal F}^{2} f_1}, \\
    C_{2 3} &= - \frac{-I_1}{{\cal F} f_2 } (\ve_1 \cdot \vf_2)
= I_1 \frac{ (\vf_1 \cdot \vf_2 ) }{{\cal F}^{2} f_2}, \\
   C_{3 3} &= - \frac{ \sin \phi }{{\cal F}^{2}} \,
     \left(
     I_2 - I_1
    + \sin \phi \, \left(
       I_2 \frac{f_2}{f_1} - I_1 \frac{f_1}{f_2} \right)
     \right).
  \end{split}
\end{equation}
At this stage we can write the final expression for the Coriolis Hamiltonian
(\ref{eq:h-cor-1}),
\begin{multline}
  \label{eq:cor-ham-f}
  H_{cor} = -i \, \frac{L_3}{{\cal F}^{2}} 
 \biggl(
 \cos \phi \,
 \left(
I_2 f_2 \, \frac{\partial }{\partial f_1}
 - I_1 f_1 \, \frac{\partial }{\partial f_2} 
\right) 
\\
+ \left( I_2 - I_1 + \sin \phi \, \left[
I_2 \frac{f_2 }{f_1 }  - I_1 \frac{f_1 }{f_2 } \right] \right)
 \, \sin \phi\,  \frac{\partial }{\partial \cos \phi} 
+ (I_1 - I_2)\, \cot \phi
\biggr).
\end{multline}
The rigid-body limit of Coriolis Hamiltonian can be obtained using
(\ref{eq:rbl}) which yields
\begin{displaymath}
  H^{(eq)}_{cor} = 0.
\end{displaymath}
Thus, in the limit of zero vibrations the vibro-rotational couplings
disappear.
This fact is the inherent property of the Eckart BF which makes it so popular
in molecular physics.

The expression (\ref{eq:cor-ham-f}) for the Coriolis Hamiltonian simplifies
when the equilibrium inertia moments are equal
$I=I_1 = I_2$,
\begin{equation}
  \label{eq:cor-ham-eq-i}
  H_{cor} = \frac{-i I}{{\cal F}^{2}}\, L_3
 \biggl(
 \cos \phi \,
 \left(
  {f_2} \, \frac{\partial }{\partial f_1}
 -{f_1} \, \frac{\partial }{\partial f_2} 
\right) 
+ \left( 
\frac{f_2 }{f_1 }  - \frac{f_1 }{f_2 } \right)
 \, (\sin \phi)^{2} \,  \frac{\partial }{\partial \cos \phi}
\biggr).  
\end{equation}


\section{The rotational Hamiltonian}
\label{sec:rotat-hamiltonian}

The rotational Hamiltonian expresses as \cite{meremianin03:_phys_rep,mylap_eng}
\begin{equation}
  \label{eq:rot-ham-1}
  H_{rot} = \frac{1}{2} \sum_{i,j=1}^{3} {\cal I}_{ij} L_i L_j,
\end{equation}
where, as above, $L_i = (\ve_i \cdot \vcl)$ denotes the $i$-th component of
the total angular momentum operator in BF and
${\cal I}$ is the effective inverse inertia tensor 
\cite{meremianin03:_phys_rep,mylap_eng}.
This tensor is symmetric
\begin{displaymath}
  \label{eq:cal-i-1}
  {\cal I}_{ij} = {\cal I}_{ji}.
\end{displaymath}
The general expression for the tensor ${\cal I}$ for the Eckart frame has been
derived in \cite{meremianin04:_eckart}.
However, in that paper the closed form for $H_{rot}$ in case of three
particles was not given.
 The more detailed analysis presented in \cite{mylap_eng} makes it
possible to derive a simpler form for the effective inertia tensor and,
thereby, for the rotational Hamiltonian.
In the three-body problem the non-zero components of the effective inertia
tensor ${\cal I}$ are \cite{mylap_eng}
\begin{equation}
  \label{eq:cal-i-2}
  \begin{split}
    {\cal I}_{11} =& 
 \frac{I_1}{{\cal F}^{2}}
\left(
1+\frac{f_1}{f_2  \sin \phi}
\right)^{2}
+ \frac{I_2 (\cot \phi)^{2} }{{\cal F}^{2}}, \\
    {\cal I}_{22} =& 
 \frac{I_2}{{\cal F}^{2}}
\left(
1+\frac{f_2}{f_1  \sin \phi}
\right)^{2}
+ \frac{I_1 (\cot \phi)^{2} }{{\cal F}^{2}}, \\
    {\cal I}_{12} =& 
 \frac{\cos \phi}{ ( {\cal F\, \sin \phi} )^{2}}
   \left(
     (I_1 + I_2) \sin \phi + I_1 \frac{f_1}{f_2}
   + I_2 \frac{f_2}{f_1}
   \right), \\
    {\cal I}_{33} =& \frac{I_1 + I_2 }{{\cal F}^{2}}.
  \end{split}
\end{equation}
The rigid-body limit of the above expressions can be easily
obtained noting eqs.~(\ref{eq:rbl}).
This leads to the well-known expression for the rotational Hamiltonian
for a rotating planar rigid system
\begin{displaymath}
H^{(eq)}_{rot} = 
 \frac{L_1^{2}}{2 I_1} + \frac{L_2^{2}}{2 I_2} 
+ \frac{L_3^{3}}{2 (I_1+I_2) }.
\end{displaymath}

The equations  (\ref{eq:cal-i-2}) simplify in practically important case
when two equilibrium moments of inertia are equal $I =I_1 = I_2$,
\begin{equation}
  \label{eq:cal-i-equal}
  \begin{split}
    {\cal I}_{11} =& \frac{I }{ ( f_2 \sin \phi)^{2}}, \quad
    {\cal I}_{22} = \frac{I }{ (f_1 \sin \phi)^{2}}, \\
    {\cal I}_{12} =& \frac{I \cos \phi}{f_1 f_2\, (\sin \phi)^{2}}, \quad
    {\cal I}_{33} = \frac{2 I}{ {\cal F}^{2}}. \\
  \end{split}
\end{equation}
Thus, the rotational Hamiltonian can be written as
\begin{equation}
  \label{eq:rot-ham-eq-i}
  H_{rot} 
= \frac{I}{2 (\sin \phi)^{2}} 
\left(
 \frac{L_1^{2}}{f_2^{2}} + \frac{L_2^{2}}{ f_1^{2}} 
+ \frac{\cos \phi}{f_1 f_2} L_1 L_2
\right)
+ \frac{I\, L^{2}_3}{{\cal F}^{2}}.
\end{equation}
This particularly simple expression has not yet been given in literature.

\section{Conclusion}
\label{sec:conclusion}

In the present paper the vibro-rotational decomposition of the
three-body Hamiltonian has been analysed  for the body-frame defined by the
Eckart condition (\ref{eq:econd-jacobi}).
The advantage of Eckart frame is that it minimises the vibro-rotational couplings
(i.e. Coriolis part of the total Hamiltonian) for small amplitude
vibrations.
The geometrical explanation of the second Eckart condition
(\ref{eq:econd-jacobi}) has been given in Sec.~\ref{sec:keo-jacobi}, see
Fig.~\ref{fig:eckart-areas}.

The derived expressions for the Coriolis (see eq.~(\ref{eq:cor-ham-f}) of
Sec.~\ref{sec:coriolis-hamiltonian}) and rotational Hamiltonians
(eqs.~(\ref{eq:rot-ham-1}), (\ref{eq:cal-i-2}) of
Sec.~\ref{sec:rotat-hamiltonian}) are written in terms of Eckart internal
coordinates which are connected (Appendix~\ref{sec:conn-eckart-jacobi}) to
conventional Jacobi-bond coordinates widely used in three-body problem
\cite{smirnov_shitikova77:_k_harm,littlejohn97:_gauge}.
The expression for the vibrational Hamiltonian in terms of Eckart variables is
given by eq.~(\ref{eq:h-vib}).
The use of the Eckart internal variables has made it possible to simplify the
expressions for the vibro-rotational Hamiltonians.

It turns out that in the case of equal equilibrium inertia moments the
Coriolis and rotational Hamiltonians can be written in particularly simple 
form, see eqs.~(\ref{eq:cor-ham-eq-i}), (\ref{eq:rot-ham-eq-i}), and
(\ref{eq:i1eqi2-2}), (\ref{eq:cal-f-i1eqi2}).
Below the sum of Coriolis and rotational
Hamiltonians is written explicitly in terms of mass-scaled Jacobi-bond
coordinates,
\begin{multline}
  \label{eq:cor-rot}
  H_{cor} + H_{rot}
= \frac{-i }{ F}\, L_3
 \left[
 \cos \theta \,
 \left(
  {r_2}  \frac{\partial }{\partial r_1}
 -{r_1}  \frac{\partial }{\partial r_2} 
\right) 
+ \left( 
\frac{r_2 }{r_1 }  - \frac{r_1 }{r_2 } \right)
 \, (\sin \theta)^{2} \,  \frac{\partial }{\partial \cos \theta}
\right] \\
+ \frac{1}{2 (\sin \theta)^{2}} 
\left(
 \frac{L_1^{2}}{r_2^{2}} + \frac{L_2^{2}}{ r_1^{2}} 
+ \frac{\cos \theta}{r_1 r_2} L_1 L_2
\right)
+ \frac{ L^{2}_3}{ F},
\end{multline}
where the parameter $F$ is defined by
\begin{displaymath}
  F = r_1^{2} + r_2^{2} + 2 r_1 r_2 \sin \theta.
\end{displaymath}
As one can easily see, the Coriolis term as well as the non-diagonal term of
the rotational Hamiltonian vanish in the above expression in the limit
$r_1 \to r_2$, $\theta \to \pi/2$ which is the equilibrium configuration with
two equal principal inertia moments.
At the same time, the rotational Hamiltonian becomes that of a rotating rigid
body.

For the sake of completeness we present here also the expression for the
vibrational kinetic energy in terms of mass-scaled Jacobi-bond coordinates,
\begin{equation}
  \label{eq:keo-vib-jbl}
  H_0 = -\sum_{i=1}^{2} \frac{1}{2 r_i^{2}} \frac{\partial}{\partial r_i}
r_i^{2} \frac{\partial}{\partial r_i}
- \frac{1}{2}
\left( 
 \frac{1}{r_1^{2}} + \frac{1}{r_2^{2}} 
 \right)
\frac{\partial}{\partial \cos \theta} (\sin \theta)^{2}
 \frac{\partial}{\partial \cos \theta}
+ U (r_1, r_2, \cos \theta),
\end{equation}
 where $U$ is the potential energy.

\appendix

\section{Relations between Eckart and Jacobi-bond coordinates}
\label{sec:conn-eckart-jacobi}

Below the connections between the set of Eckart and Jacobi internal coordinates
are presented.
Such connections follow from the definition of Eckart vectors
(\ref{eq:def-f-vect}) and the properties of the principal inertia axes.
Omitting details of somewhat cumbersome computations we present here only the
final results,
\begin{equation}
  \label{eq:f1sq}
  f_1^{2} 
= 
\frac{I_2}{I_1 - I_2}
\left( r_1^{2} (I_1 - \rho_1^{2}) + r_2^{2} (I_1 - \rho_2^{2}) 
- 2 (\vvr_1 \cdot \vvr_2 )\, ( \brho_1 \cdot \brho_2 )
\right). 
\end{equation}
For the vector $\vf_2$ one has
\begin{equation}
  \label{eq:f2sq}
  f_2^{2} 
= 
\frac{-I_1}{I_1 - I_2}
\left(
 r_1^{2} (I_2 - \rho_1^{2}) + r_2^{2} (I_2 - \rho_2^{2}) 
- 2 (\vvr_1 \cdot \vvr_2 )\, ( \brho_1 \cdot \brho_2 )
\right),
\end{equation}
The scalar product of Eckart vectors expresses as
\begin{equation}
  \label{eq:f1f2}
  (\vf_1 \cdot \vf_2)
= \frac{| \brho_1 \times \brho_2 |}{I_1 - I_2}
\left(
 (\brho_1 \cdot \brho_2)\, (r_1^{2}-r_2^{2}) - 
(\vvr_1 \cdot \vvr_2)\, (\rho_1^{2} - \rho_2^{2}) 
\right).
\end{equation}
As is seen, the scalar product $(\vf_1 \cdot \vf_2)$ vanishes in the
rigid-body limit when the vibration amplitudes tend to zero.
The analysis of the rigid body limit of parameters $f_1$ and $f_2$ is slightly
more complicated.
Namely, in the rigid-body limit we have $\vvr_1 \to \brho_1$ and 
$\vvr_2 \to \brho_2$ which, noting the definitions (\ref{eq:def-f-vect}),
leads to
\begin{equation}
  \label{eq:rbl-f1}
  \left(f_1^{2} \right)_{eq} = x_1^{2} \rho_1^{2} + x_2^{2} \rho_2^{2}
+ 2 x_1 x_2 (\brho_1 \cdot \brho_2). 
\end{equation}
Now we re-write $\rho$'s in terms of their coordinates according to
(\ref{eq:brho-comp}). 
This yields
\begin{equation}
  \label{eq:rbl-f1-2}
   \left(f_1^{2} \right)_{eq} = \left( x_1^{2} + x_2^{2} \right)^{2}
+ (x_1 y_1 + x_2 y_2)^{2} = I_2^{2}.
\end{equation}
Here, in the course of derivations we have utilized eqs.~(\ref{eq:bf-2}) and
(\ref{eq:def-I}).

Thus, in the limit of zero vibrations one has
\begin{equation}
  \label{eq:rbl}
  \begin{split}
    (f_1)_{eq} &= I_2, \quad
    (f_2)_{eq} = I_1, \\
    \phi_{eq} & = \frac{\pi}{2}.
  \end{split}
\end{equation}

Clearly, the above eqs.~(\ref{eq:f1sq}) -- (\ref{eq:f1f2}) are invalid when the
equilibrium principal moments of inertia are equal.
According to eq.~(\ref{eq:inertia-moms}) of the next Appendix the condition
$I_1 = I_2$ is met only if 
\begin{equation}
  \label{eq:i1eqi2}
  \rho_1 = \rho_2, \quad \theta_e = \pi/2,
\end{equation}
where $\theta_e$ is the mutual angle in the equilibrium configuration, see
Fig.~\ref{fig:3bd-jacobi}.
Thus, the vectors $\brho_1$ and $\brho_2$ are perpendicular and, hence, we can
choose them to define the Cartesian basis, i.e.
\begin{equation}
  \label{eq:i1eqi2-1}
  \ve_1 = \frac{\brho_1}{\rho_1},  \quad
  \ve_2 = \frac{\brho_2}{\rho_2}.
\end{equation}
As a consequence we have that the coordinates  in eq.~(\ref{eq:brho-comp})
become
\begin{displaymath}
x_1 = \rho_1, \quad y_2 = \rho_2, \quad
x_2 = y_1 =0.
\end{displaymath}
From these identities and from (\ref{eq:def-f-vect}) we obtain that at $I_1 =
I_2$ the Eckart vectors can be chosen as
\begin{equation}
  \label{eq:i1eqi2-2}
  \vf_1 = \rho \vvr_1, \quad \vf_2 = \rho \vvr_2,
\end{equation}
where $\rho = \rho_1 = \rho_2$.
The corresponding Eckart parameter (\ref{eq:2d-trace-f}) expresses as
\begin{equation}
  \label{eq:cal-f-i1eqi2}
  {\cal F}^{2} = \rho^{2} \, \bigl( r_1^{2} + r_2^{2}
+ 2 r_1 r_2 \sin \theta \bigr).
\end{equation}

\section{The equilibrium inertia moments}
\label{sec:equil-I}

The straightforward calculation of the eigenvalues of the inertia
tensor leads to the following relations for the principal inertia moments of
the equilibrium configuration
\begin{equation}
  \label{eq:inertia-moms}
  \begin{split}
    I_1 &= \frac{\rho_1^{2}+\rho_2^{2}}{2} + \frac{1}{2}\sqrt{
      (\rho_1^{2} - \rho_2^{2})^{2} +  (2 \rho_1 \rho_2 \cos  \theta_e)^{2}}, \\
  I_2 &=  \frac{\rho_1^{2}+\rho_2^{2}}{2} - \frac{1}{2}\sqrt{
      (\rho_1^{2}-\rho_2^{2})^{2} + (2 \rho_1 \rho_2 \cos \theta_e )^{2}}.
  \end{split}
\end{equation}
We remind that $\rho_{1,2}$ are mass-scaled Jacobi vectors.
To obtain expressions in terms of conventional Jacobi vectors one should apply
the replacements $\rho_{1,2} \to \rho_{1,2} \, \sqrt\mu_{1,2}$
to the above equations (\ref{eq:inertia-moms}).

As is seen from (\ref{eq:inertia-moms}) the principal inertia moments are
equal if Jacobi vectors are perpendicular (i.e. when $\theta_e = \pi/2$) and 
their lengths satisfy the equation
\begin{equation}
  \label{eq:i12-eq}
  \mu_1 \rho_1^{2} = \mu_2 \rho_2^{2}.
\end{equation}


\end{document}